\documentclass[%
 reprint,
 amsmath,amssymb,
 aps,
]{revtex4-2}

\usepackage{graphicx}
\usepackage{dcolumn}
\usepackage{bm}

\begin{document}

\preprint{APS/123-QED}
\title{Encapsulation-Induced Alignment in Endofullerenes}

\author{Jonathan Smucker}
\email{jonathan.smucker@stonybrook.edu}
\affiliation{%
Department of Physics and Astronomy, Stony Brook University, Stony Brook
}%

\author{Jesus P\'erez-R\'ios}%
\affiliation{%
Department of Physics and Astronomy, Stony Brook University, Stony Brook
}%

\date{\today}

\begin{abstract}
Methods for creating endofullerenes have been steadily improving since their discovery, allowing for new types of endofullerenes to be created in larger numbers. When a molecule is trapped in a fullerene, the fullerene creates a harmonic trapping potential that leaves most of the fundamental properties of the internal molecule intact. The fullerene cage does create a preferred axis for the internal molecule, which we refer to as the encapsulation-induced alignment of the molecule. We explore the alignment of AlF and N$_2$ inside of C$_{60}$ by first computing the interaction between the internal molecule and the fullerene cage using {\it ab initio} electronic structure methods. Our results show that the internal molecules are found to be strongly aligned despite finding that all the calculated spectroscopic constants are relatively unaffected by the fullerene cage.
\end{abstract}

\maketitle


\section{Introduction}
When first discovered, Buckminster fullerene (C$_{60}$) was thought to be of niche interest to astrophysics. Methods for producing C$_{60}$ in a laboratory were rapidly developed and fullerenes were discovered to have many applications in physics, chemistry, astronomy and material science \cite{krotoC60Rev}. Rapidly expanding studies of fullerenes quickly discovered that atoms and molecules could be trapped inside the hollow carbon cage \cite{MetalEndoRev, EndoRev, FirstEndo}, these complexes were named endohedral fullerenes or just endofullerenes. When the internal atom or molecule is light enough, the rotation and vibration of the internal molecule becomes quantized and cannot be treated classically \cite{LightMoleculesEndo}. When the internal atom or molecule of an endofullerene is sufficiently heavy, it maintains many of its individual properties \cite{MetalEndoRev}, similar to a molecule or atom inside a carbon nanotube. The internal molecule or atom is therefore protected from outside collisions while effectively remaining unperturbed. These properties have drawn attention from researchers due to possible applications in medical science, photovoltaics, quantum computing, and electronics \cite{MetalEndoRev,EndoRev,EndoQCRev}.

Thanks to experimental developments over the last 40 years, a wide range of molecules have been trapped inside fullerene cages \cite{EndoRev, LightMoleculesEndo}. Of particular interest is the development of molecular surgery, a process in which a fullerene cage is chemically pried open allowing an atom or small molecule to be placed inside before the fullerene is sealed again \cite{SurgeryRev}. Modern experimental methods allow much greater control over what type of molecular or atom is trapped inside the fullerene and allow more diverse sets of endofullerenes to be produced. With molecular endofullerenes more available than ever, more and more studies are focusing on the properties of these endofullerenes rather than their production. Specifically, many of these studies focused on the translational and rotational energy of the trapped atom or molecule \cite{AtomEndoInt,HeTrans,ye2010quantum,xu2009coupled,felker2016translation}. Some of these studies use the sum of pairwise Lennard-Jones type interactions to predict the interaction between the internal molecule with the outer fullerene wall. We instead employ a grid of single point energy calculations to obtain our potential energy surfaces. Studies of vibration are limited.

This study focuses on the alignment of molecules inside of C$_{60}$, which we refer to as the encapsulation-induced alignment. The fullerene creates a preferred axis for the internal molecule, similar to the preferred axis created by an external electric field in alignment studies of free molecules. Our study focuses on two endofullerenes: N$_2$@C$_{60}$ and AlF@C$_{60}$. We demonstrate how the fullerene cage shifts the spectroscopic constants of these internal molecules. We also briefly study the effects of the fullerene on the dipole moment of AlF. N$_2$ represents a molecule with a highly covalent bond and a short equilibrium distance. AlF has a much higher ionic character and a much longer equilibrium distance. The contrast between these two internal molecules helps to provide insight on how a wider variety of molecules might behave while trapped inside a fullerene cage.

\section{Computational Methods}

The optimized geometries and the energies of various configurations were calculated using density functional theory (DFT) as implemented in Gaussian 16 \cite{g16}. First, we performed a comparative study for the encapsulation energy of N$_2$@C$_{60}$ using different functionals and basis sets, and the results are shown in Table~\ref{tab:comp}. In terms of the geometry of the endofullerene, we notice that the outer C$_{60}$ cage is largely unchanged from method to method and largely unchanged through the introduction of the internal N$_2$ molecule. The only major difference between the optimized geometries produced by the different methods is the bond length of the internal molecule, which results in distinct encapsulation energies. 

We note that the B3LYP\cite{B3LYP} and PBEH1PBE~\cite{PBEh1PBE} functionals, which are well suited to describe covalent interactions, predict positive binding energies. However, functionals that are better suited for unbonded interactions, such as MN15~\cite{MN15}, PW6B95~\cite{PW6B95}, APFD~\cite{APFD}, and wB97XD~\cite{wB} show negative binding energies. These findings agree with previously published results on the same system~\cite{Rec2Encamp}, although these results used fixed geometries for C$_{60}$. Moreover, functionals including dispersion corrections or empirical dispersion corrections predict negative encapsulation energies with the exception of CAM-B3LYP~\cite{CAM-B3LYP}, although the encapsulation energy for CAM-B3LYP is very small. Based on the nature of the two systems under consideration: AlF@C$_{60}$ and N$_2$@C$_{60}$, we conclude that it is appropriate to use a functional with dispersion correction to accommodate weak interactions beyond the covalent interactions common in C$_{60}$. This conclusion agrees with other studies that emphasize the requirement of functionals that include dispersion corrections to accurately describe observed spectra in endofullerenes~\cite{HeEndoSpect}.


\begin{table}[h]
\setlength{\tabcolsep}{10pt}
\centering
\caption{Comparison of the encapsulation energy of N$_2$@C$_{60}$ calculated using various different methods. None of these calculations accounted for basis superposition error (BSSE).}\label{tab:comp}
\begin{tabular}{ccc}
\hline
Method & \begin{tabular}[c]{@{}c@{}}Def2TZV\\ Basis (kcal/mol)\end{tabular} & \begin{tabular}[c]{@{}c@{}}6-31G**\\ Basis (kcal/mol)\end{tabular} \\ \hline
\begin{tabular}[c]{@{}c@{}}B3LYP-D3\end{tabular} & -15.87 & -14.63 \\ \hline
B3LYP & 6.70 & 8.07 \\ \hline
CAM-B3LYP & 0.29 & 1.76 \\ \hline
MN15 & -23.68 & -19.36 \\ \hline
PW6B95 & -11.85 & -12.12 \\ \hline
PW6B95D3 & -20.63 & -20.85 \\ \hline
PBEh1PBE & 0.42 & 1.24 \\ \hline
wB97XD & -12.12 & -12.04 \\ \hline
APFD & -18.79 & -18.04 \\ \hline \hline
\end{tabular}
\end{table}

\begin{figure}
\begin{center}
\includegraphics[width=8.5 cm]{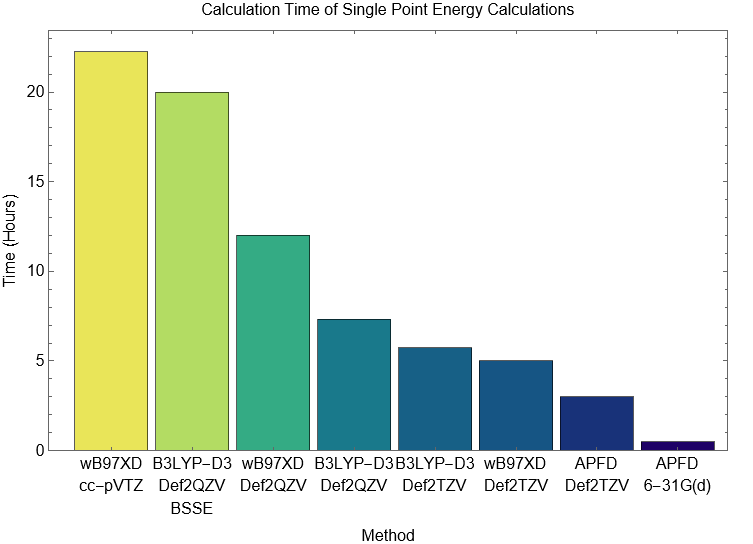}
\caption{Comparison of the calculation times for single point calculations using various methods. Only the second method (shown in green) accounts for the basis set superposition error using the counterpoise method \cite{CP1,CP2}. \label{figComp}}
\end{center}
\end{figure}

From studies comparing experimental spectroscopic data on He@C$_{60}$~\cite{HeEndoSpect}, it has been shown that the wB97XV functional leads to the most accurate results, followed by B3LYP-D3. Fueled by the results displayed in Table~\ref{tab:comp} and following that recommendation, we tested the computational cost of the of the wB97XV~\cite{wB} functional, the B3LYP functional with GD3BJ empirical dispersion (B3LYP-D3) and the APFD functional. These results are displayed in Fig.~\ref{figComp}. The computational cost of the wB97XV method is comparable to the cost of the B3LYP-D3 method when using the Def2TZV basis set. When using the Def2QZV basis set, the cost of the wB97XV functional surpassed the cost of the B3LYP-D3 method significantly. For this reason, we used the B3LYP-D3 method as a compromise between precision and computational cost. This is the method that we used for all the DFT calculations reported in this study, unless otherwise stated. All our calculations account for the basis set superposition error (BSSE) using the counterpoise method \cite{CP1,CP2} unless otherwise stated.



For the choice of basis set we ran calculations with the cc-pVnZ (n being D,T,Q,5 etc.) class of basis sets, the 6-31G** basis set and the Def2nZV class of basis sets. The Def2nZV class of basis sets are optimized for DFT calculations and significantly smaller than there cc-pVnZ counter parts. The 6-31G** basis set is small and likely inaccurate. The cc-pVnZ basis sets proved to be impractical. For N$_2$@C$_{60}$ we used the Def2QZV basis set, particularly because it was the largest basis set that was still practical in terms of computation times. For AlF@C$_{60}$ the computation times were roughly double. To reduce computational cost, we used the Def2QZV basis for the optimization and the Def2TZV basis for calculations of the energy as a function of stretching and rotation. The energies for AlF@C$_{60}$ as a function of rotation were also found to be a factor of 10 larger then for N$_2$@C$_{60}$. Due to this, the difference between the Def2QZV basis set and the Def2TZV basis set was significant for N$_2$@C$_{60}$ but not for AlF@C$_{60}$, justifying the use of the smaller basis.

\begin{figure}
\begin{center}
\includegraphics[width=8.5 cm]{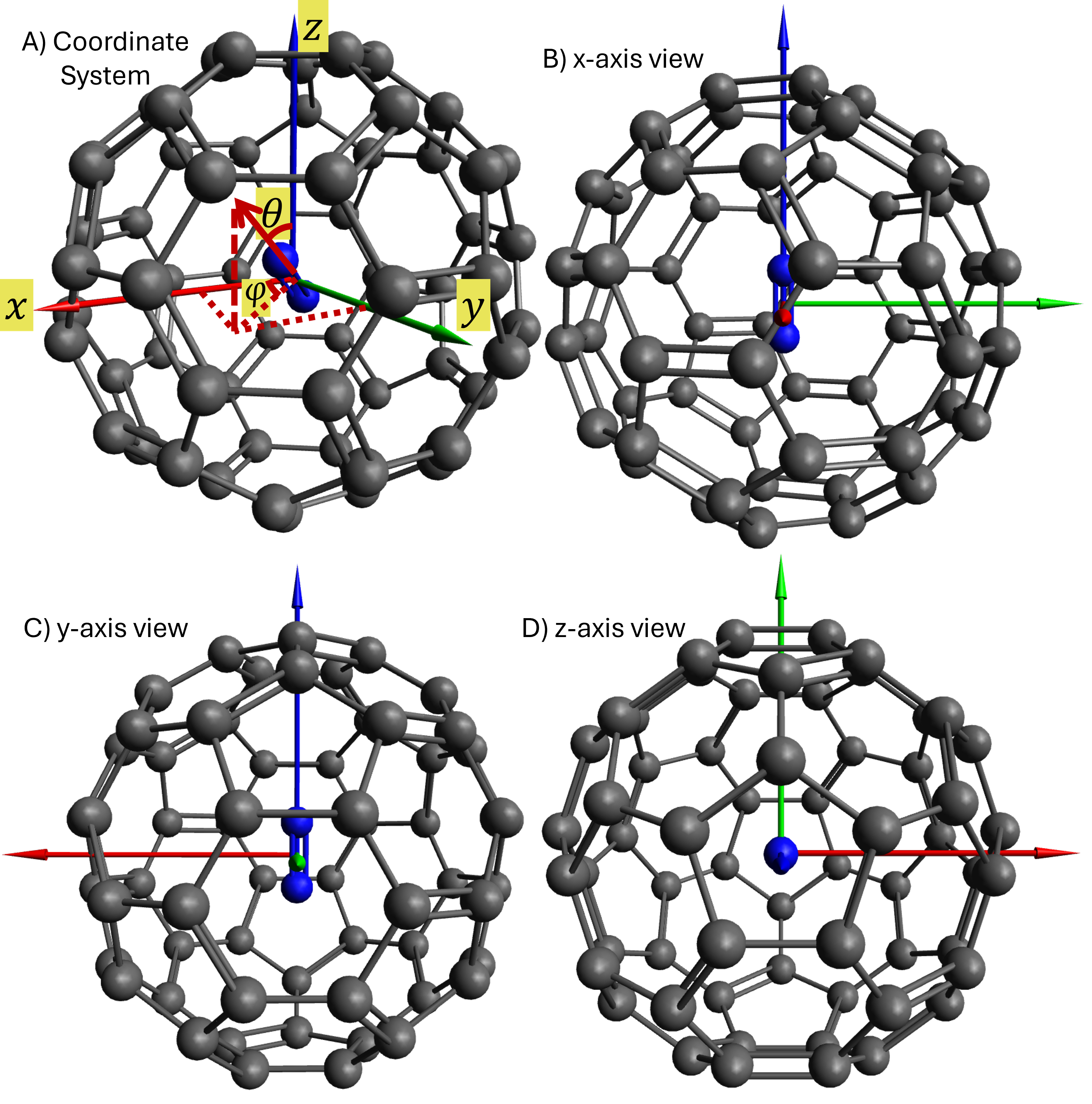}
\caption{The optimized geometry of N$_2$@C$_{60}$. Panel A shows the coordinate system used throughout this paper. Panels B, C and D show the optimized geometry of N$_2$@C$_{60}$ as viewed from the x, y and z-axes respectively \cite{Avagadro}.\label{figN2Geom}}
\end{center}
\end{figure}

\begin{figure}
\begin{center}
\includegraphics[width=8.5 cm]{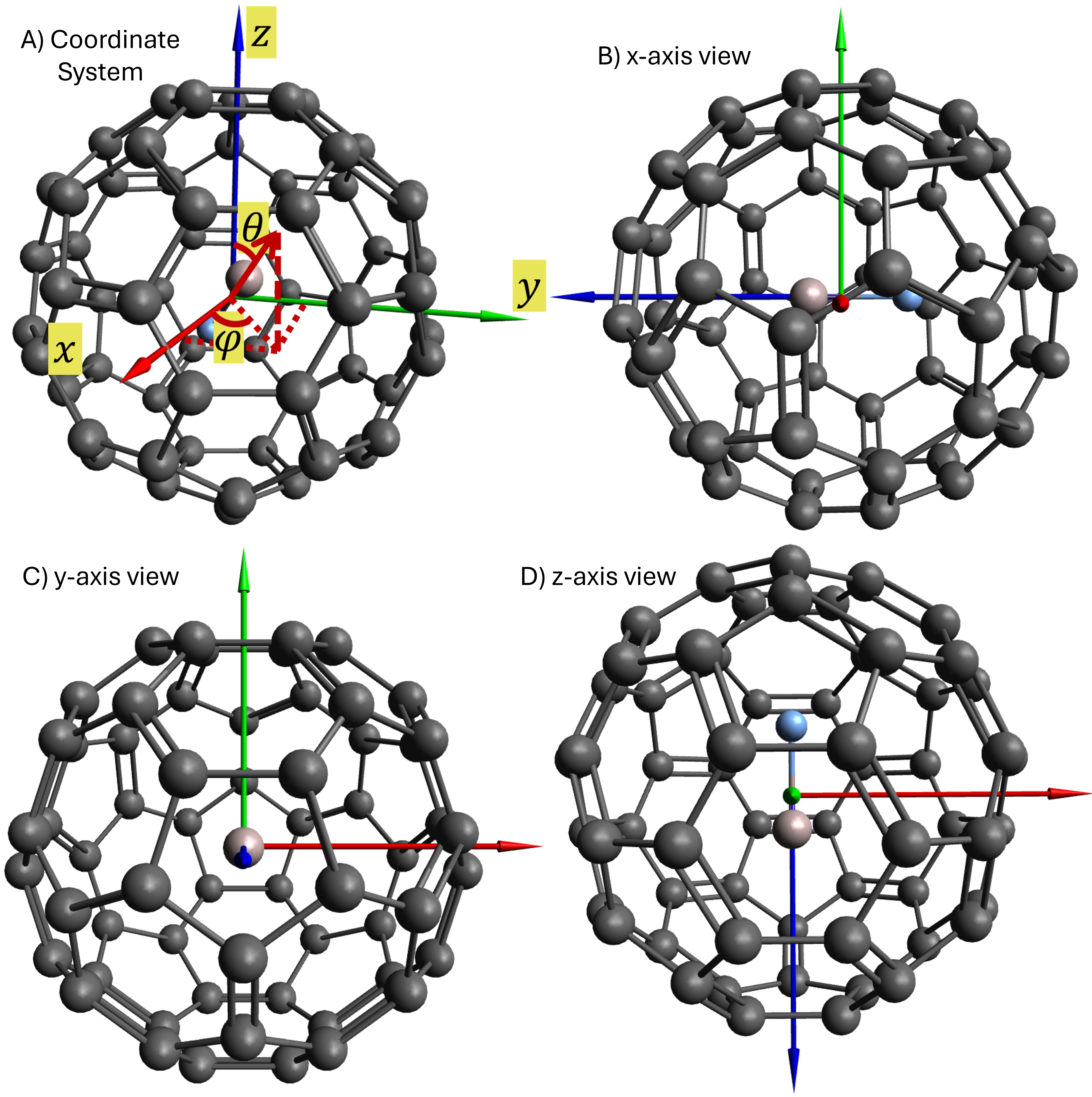}
\caption{The optimized geometry of AlF@C$_{60}$ .Panel A shows the coordinate system used throughout this paper. Panels B, C and D show the optimized geometry of AlF@C$_{60}$ as viewed from the x, y and z-axes respectively \cite{Avagadro}.\label{figAlFGeom}}
\end{center}
\end{figure}

The optimized geometry calculated for N$_2$@C$_{60}$ at the B3LYP level of theory with the Def2QZV basis set with counterpoise correction and GD3BJ dispersion included is shown in Fig.~\ref{figN2Geom}. The optimized geometry for AlF@C$_{60}$ calculated using the same method is shown in Fig.~\ref{figAlFGeom}. N$_2$ is found to be in the exact center of the fullerene. While, AlF is slightly shifted off center. The optimized geometry of both endofullerenes is shown to give D$_{5d}$ symmetry as previously predicted \cite{Rec1CompN2,Rec2Encamp,EndoBook}.

The equilibrium bond length of AlF inside the C$_{60}$ cage was found to be about 1.69 \AA~as compared to the 1.65 \AA~equilibrium bond length expected for free AlF \cite{DiatDataBase,MolSpect}. For N$_2$, the equilibrium bond length inside of the C$_{60}$ cage was found to be 1.11 \AA~compared to 1.10 \AA~ for the free molecule \cite{DiatDataBase,MolSpect}. The AlF molecule is predicted to stretch slightly more due to the cage when compared to N$_2$. This might be expected because of the larger bond length, making the nuclei closer to the cage, and due to the dipole moment causing stronger long-range reactions. However, using the same methods, the equilibrium distance of free AlF is calculated to be 1.74 \AA. Since this is larger than the experimentally determined bond length of free AlF (1.65 \AA) \cite{DiatDataBase} and larger than the bond length predicted for AlF inside of a fullerene (1.69 \AA), these disagreements are likely the result of uncertainties in our calculations. The change in bond length for N$_2$ is even smaller and therefore is even more likely to be insignificant.

\section{Alignment and Orientation}

When a molecule is placed inside a fullerene, assuming that its center of mass does not move (neglecting translational degrees of freedom), the interaction energy could depend on the orientation of the molecule. If that is the case, the anisotropy of the interaction will mix up the rotational states, causing alignment and orientation effects.



Using the results from the DFT calculations, we can calculate various spectroscopic properties of the internal molecule and calculate the degree of alignment and orientation of the molecule. Couplings between rotational states can be calculated given a specific bond length $R$ for the internal molecule by expanding in the unperturbed rotational basis:
\begin{equation}
\label{eq1}
    \Psi(\theta,\phi)=\sum_{m=-l}^{l}\sum_l^\infty c_{lm} Y_l^m(\theta,\phi),
\end{equation}
where $\theta$ is the angle from the z-axis and $\phi$ is the angle from the x-axis (the typical definitions for spherical coordinate, see Fig.~\ref{figAlFGeom}a and of Fig.~\ref{figN2Geom}a). 

The Hamiltonian of the molecule is given by $\hat{H}=\hat{H}_{rot} + \hat{H}_{c}$, where $\hat{H}_{rot}$ represents the rigid rotor Hamiltonian, which is diagonal in the basis given by Eq.~(\ref{eq1}), $H_{rot,l,m,l',m'}=B_e l(l+1)\delta_{ll'}\delta_{mm'}$. Here, $B_e$ is the rotational constant of the ground state of the free molecule. It is worth noticing that the rotational constant of the free molecule is roughly equal to the rotational constant predicted for the internal molecule. $\hat{H}_{c}$ represents the interaction energy between the molecule and the cage, and its matrix elements are given by:
\begin{equation}
    H_{c,l,m,l',m,}=\langle Y_l^{*m}(\theta,\phi)|V(R,\theta,\phi)| Y_l'^{m'}(\theta,\phi)\rangle,
\end{equation}
which is non-diagonal. The potential $V(R,\theta,\phi)$, is produced by interpolating a grid of single point energy calculations. As previously stated, these calculations were done at the B3LYP level of theory with the Def2QZV basis set for N$_2$@C$_{60}$ and with the Def2TZV basis set for AlF@C$_{60}$. The calculations for both molecules include counterpoise corrections and GD3BJ dispersion. The energy as the molecule stretches is significantly greater than the energy as the molecule rotates for both AlF and N$_2$. This allows these two degrees of freedom to be treated separately and for the internal molecule to be treated as a rigid rotor. 

\begin{figure}
\begin{center}
\includegraphics[width=8.5 cm]{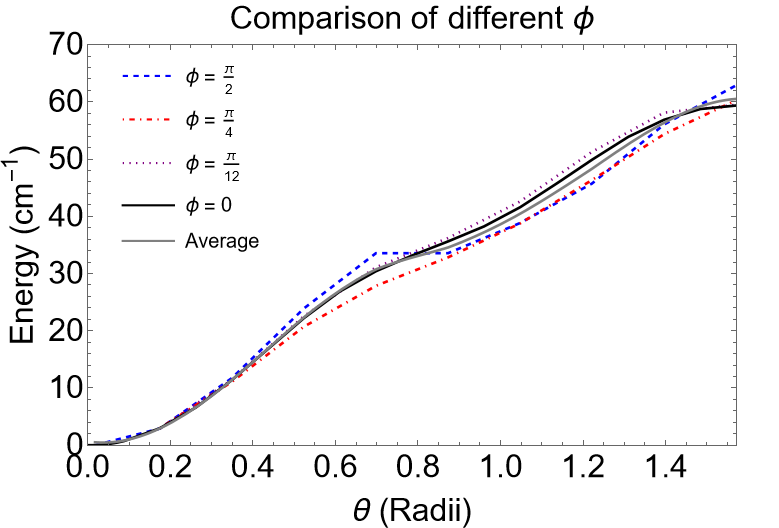}
\caption{The energy of N$_2$@C$_{60}$ as the internal N$_2$ molecule rotates along the angle $\theta$. This energy is shown for various angles $\phi$. The gray lines represents the average of all the angles $\phi$. \label{figPhi}}
\end{center}
\end{figure}

Fig.~\ref{figPhi} shows a comparison of the energy as a function of the angle $\theta$ for various angles $\phi$. The difference between these curves never exceeds $10.4\%$, so it is accurate to treat the potential as independent of $\phi$. This result implies that $V(R,\theta)$ is enough to describe the interaction of a molecule in a fullerene, and hence 
\begin{equation}
\label{eq3}
H_{l,l'}=B_e l(l+1)\delta_{ll'}+\langle Y_l^{0}(\theta,\phi)|V(R,\theta,0)| Y_{l'}^{0}(\theta,\phi)\rangle.
\end{equation}
In other words, states with different values of $m$ will not mix and only rotational states with $m=0$ contribute to the rotational ground state.


\begin{figure}
\begin{center}
\includegraphics[width=8.5 cm]{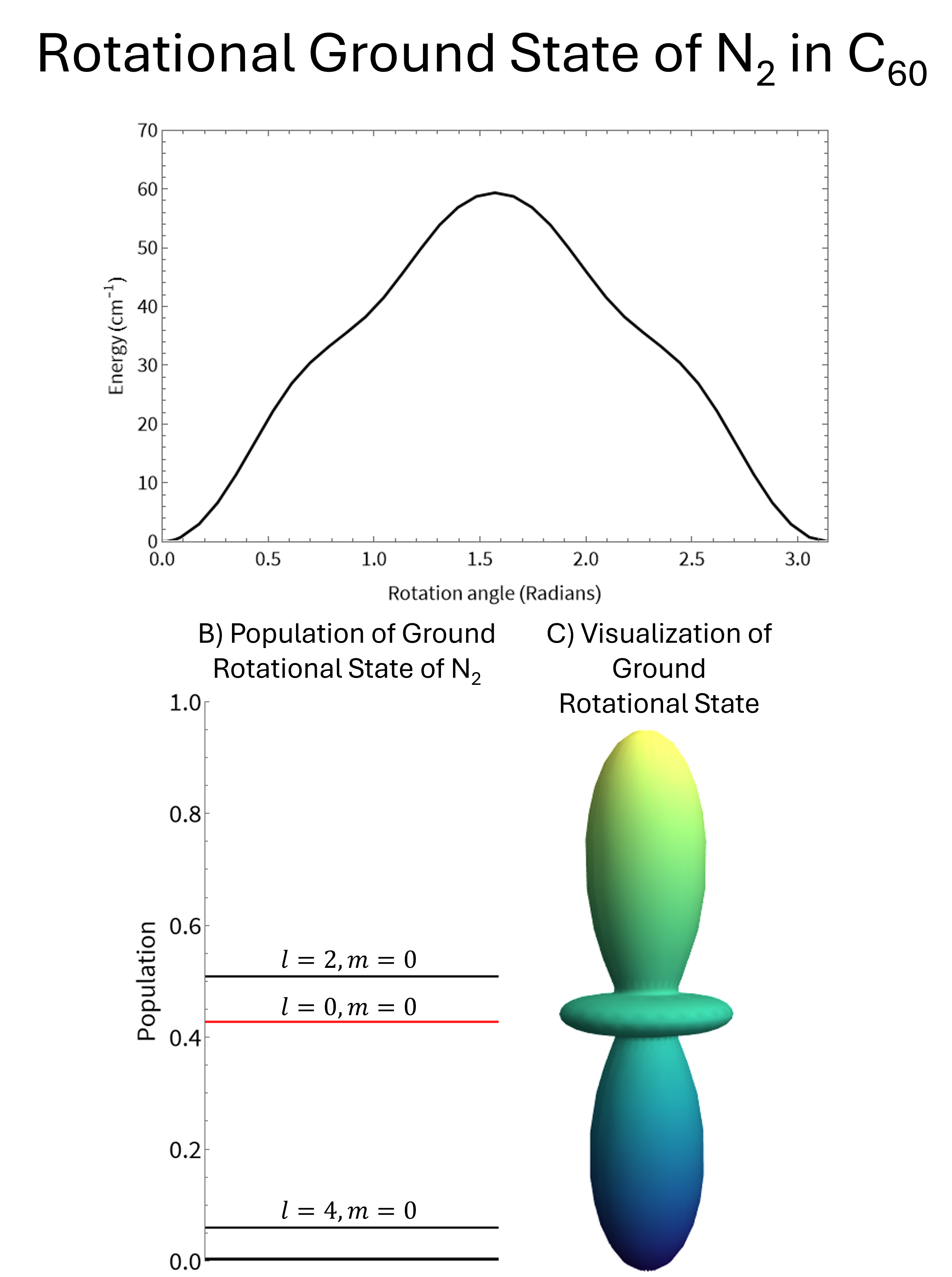}
\caption{Mixing of the rotational states of N$_2$ due to the interaction with the C$_{60}$ cage. Panel A shows the populations of the rotational states of a free N$_2$ molecule in the new ground rotational state. Panel B shows the new ground state as a spherical plot. Panel C shows the energy as a function of rotation of the N$_2$ molecule inside the C$_{60}$ cage. \label{figN2Rotor}}
\end{center}
\end{figure}

Assuming that the internal molecules are rigid rotors, the diagonalization of Eq.~(\ref{eq3}) reveals the rotational states of the molecule inside the fullerene, which can be used to analyze the changes to the rotational states caused by the interaction with the fullerene cage. These results for N$_2$@C$_{60}$ are shown in Fig.~\ref{figN2Rotor}. Due to the symmetry of N$_2$, the potential is symmetric around $\theta= 90^{\circ}$. This results in the molecule being strongly aligned but not very strongly oriented. A strong alignment corresponds to high populations in states with even values of $l$ since alignment is calculated as: $\langle \cos^2(\theta) \rangle$. Because the potential is relatively weak, the ground state is mainly composed of the lowest-lying even $l$ state, which is the $l=2$ state. Therefore, the ground state appears as a slightly altered version of the $l=2$ spherical harmonic, shown in panel B of Fig.~\ref{figN2Rotor}. This weak mixing means that the new ground state can be described accurately by only accounting for a few unperturbed rotational states. Our calculations of the perturbed rotational ground state accounted for eleven unperturbed rotational states. However, sufficient convergence was found by accounting for as little as five states, as the populations of the higher $l$ values were found to rapidly approach zero. The alignment of the new ground state was calculated to be 0.82 as opposed to the orientation ($\langle \cos(\theta) \rangle$) of the new ground state which was calculated to be 0.0053. The small orientation should not come as a surprise, as the N$_2$ molecule is homonuclear.

\begin{figure}
\begin{center}
\includegraphics[width=7.5 cm]{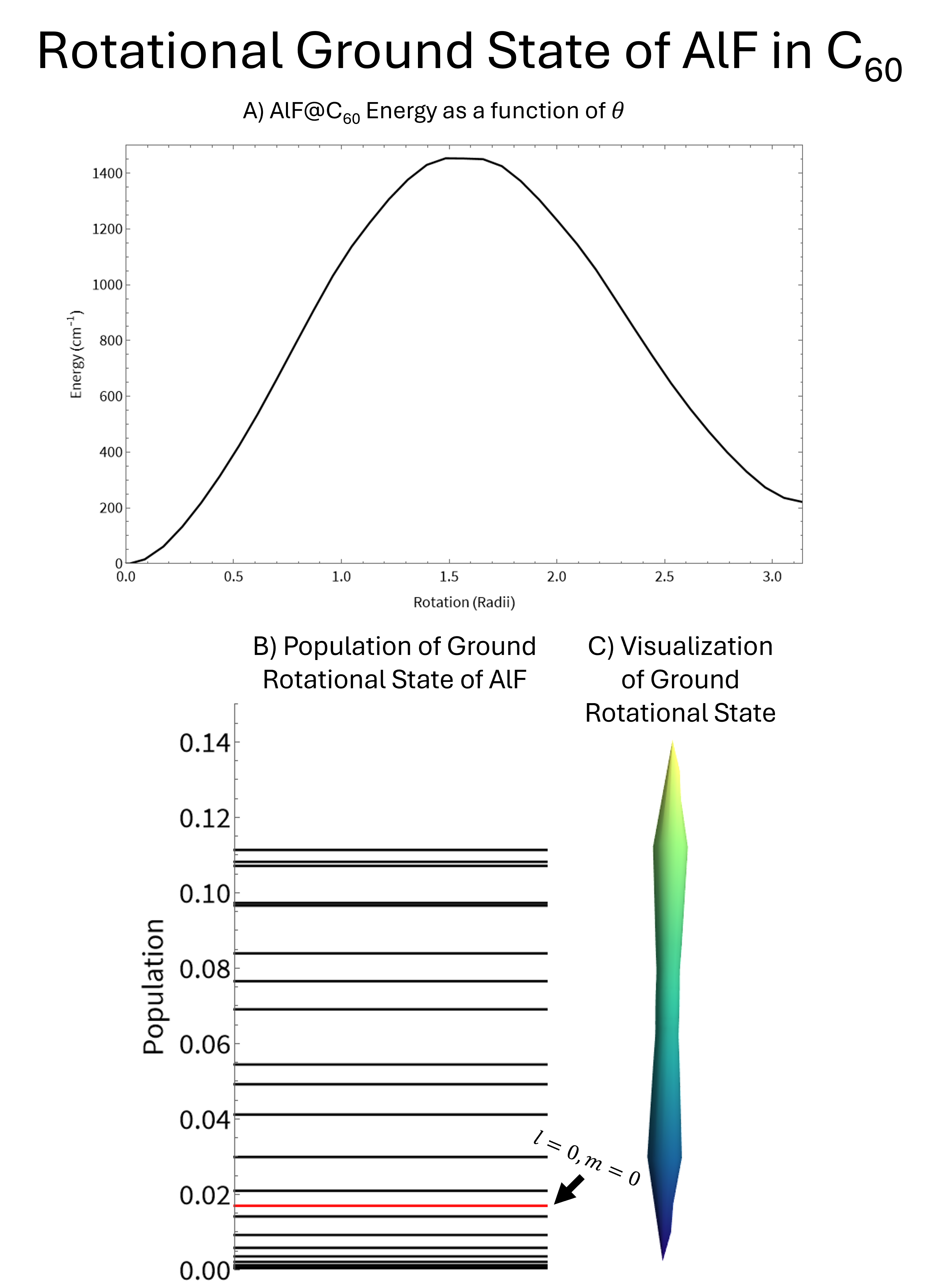}
\caption{Mixing of the rotational states of AlF due to the interaction with the C$_{60}$ cage. Panel A shows the populations of the rotational states of a free AlF molecule in the new ground rotational state. The new ground state is thoroughly mixed with no clear preference for even or odd values of $l$. Panel B shows the new ground state as a spherical plot. Panel C shows the energy as a function of rotation of the AlF molecule inside the C$_{60}$ cage.\label{figAlFRotor}}
\end{center}
\end{figure}

The potential of AlF as it rotates inside the C$_{60}$ molecule, shown in panel C of Fig.~\ref{figAlFRotor}, has a few important differences from N$_2$. The potential for AlF is symmetric around a 180$^{\circ}$ rotation. The energy at 180$^{\circ}$ is higher than the energy at the optimal geometry, as the Al-C and F-C interactions are very different. This forces the AlF molecule to be strongly oriented and strongly aligned. The alignment of the new ground state was calculated to be 0.98 and the orientation of the new ground state was calculated to be 0.99. The increased alignment is due to the increased strength of the potential. Compared to N$_2$@C$_{60}$, the energy as a function of rotation is approximately ten times stronger for AlF@C$_{60}$. This results in the perturbed rotational ground state becoming a heavy mix of a wide variety of unperturbed rotational states. To calculate the new rotational ground state, we needed to account for approximately 27 unperturbed states in order to find sufficient convergence, compared to the five needed for N$_2$. Since the molecule is aligned and oriented, there is no strong preference for states with an even value of $l$ like what we predicted for N$_2$. The strong alignment causes the visualization of the new ground state, shown in panel B of Fig.~\ref{figAlFRotor}, to be very long and thin. The many unperturbed states represented within the new ground state cause the visualization to not closely resemble any one spherical harmonic.

We find that the change in energy as the internal molecules stretch is significantly greater than the change in energy as the internal molecule rotates for both endofullerenes studied. Panel C of Fig.~\ref{figAlFRotor} shows that the energy as a function of rotation for AlF inside of C$_{60}$ is on the order of $10^3$ cm$^{-1}$. Compared to the energy as a function of stretching, shown in panel B of Fig.~\ref{figAlFRotStr}, which is on the order of $10^4$ cm$^{-1}$. This energy difference is even greater in the case of N$_2$@C$_{60}$ due to the tight bond between the two nitrogen molecules. This difference in energy allows for the rotational and vibrational degrees of freedom to be treated separately, as we have done, without significant loss in accuracy.

\section{Spectroscopic Constants}
\begin{figure}
\begin{center}
\includegraphics[width=7.5 cm]{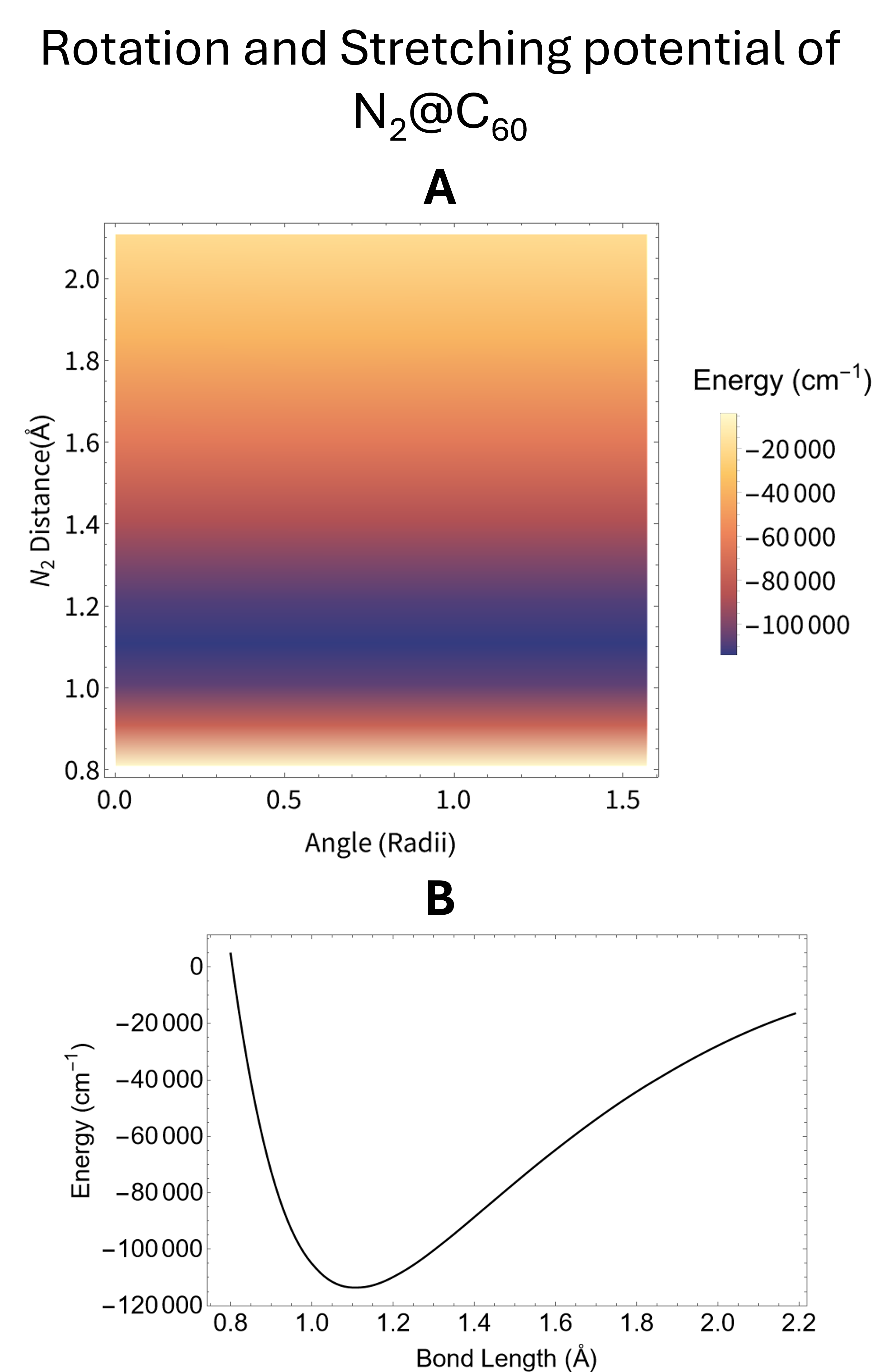}
\caption{The potential for the N$_2$ molecule within a fullerene. Panel A shows the potential energy as a function of both rotation and stretching calculated using DFT. Panel B shows the rotational average of the potential shown in panel A. \label{figN2RotStr}}
\end{center}
\end{figure}

\begin{figure}
\begin{center}
\includegraphics[width=7.5 cm]{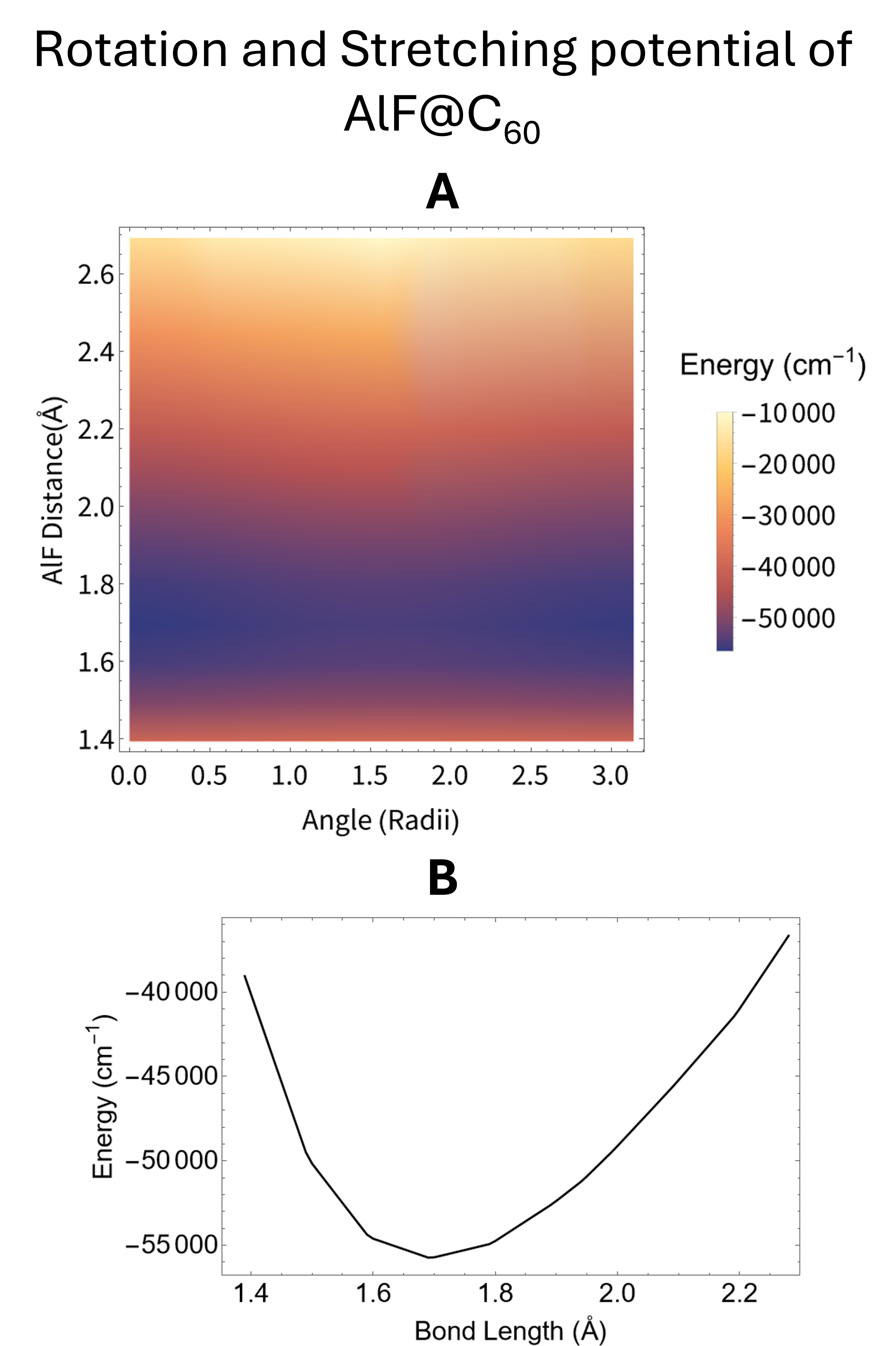}
\caption{The potential for the AlF molecule within a fullerene. Panel A shows the potential energy as a function of both rotation and stretching calculated using DFT. Panel B shows the rotational average of the potential shown in panel A.\label{figAlFRotStr}}
\end{center}
\end{figure}

The potential, $V(r,\theta,\phi)$, as a function of both rotation and stretching is shown in panel A of Figs.~\ref{figN2RotStr} and Fig.~\ref{figAlFRotStr} for N$_2$@C$_{60}$ and AlF@C$_{60}$, respectively. On both of these plots the energy appears to change little as the rotation angle changes. This difference in energy allows for the rotational and vibrational degrees of freedom to be treated separately without a significant loss in accuracy. Therefore, it is possible to introduce the rationally averaged potential as: 

\begin{align}
    \widetilde{V(r)}=\langle V(r,\theta,\phi) \rangle = \int_0^{2\pi} d\phi \int_0^{\pi} d\theta~Y_l^{m*}(\theta,\phi)~\nonumber\\ \times V(r,\theta,\phi) ~Y_l^m(\theta,\phi) \sin{\theta},
\end{align}
where $r$ is the internuclear separation of the internal molecule. Panel B of Fig.~\ref{figN2RotStr} and Fig.~\ref{figAlFRotStr} shows the rotationally averaged potential ,$\widetilde{V(r)}$, for N$_2$@C$_{60}$ and AlF@C$_{60}$, respectively. We notice that for both systems the interaction potential resembles a Morse potential, characteristic of covalent systems, especially N$_2$@C$_{60}$. 

Assuming that the energy as a function of nuclear separation is described by a Morse potential, we find 
\begin{equation}
\label{eq5}
\widetilde{V(r)} = D_e(1-e^{a(R-R_e)})^2-D_e,
\end{equation}
where $D_e$ is the dissociation energy, $R_e$ is the equilibrium distance given by the DFT calculations and $a$ is a fit parameter related with anharmonicity of the potential. To obtain $a$, we adjusted $D_e$ until the R-squared value of the fit was maximized. This technique introduces a source of error, but this error should be smaller than the errors already introduced through the use of DFT. This technique has the benefit of allowing us to use existing theories done for Morse potentials particularly used to describe diatomic molecules~\cite{demtrodertext}. 

\begin{table*}[]
\setlength{\tabcolsep}{12pt}
\large
\begin{center}
\caption{Constants calculated for AlF and N$_2$ inside of C$_{60}$ using our methods compared to known values for the ground states of free AlF and N$_2$ \cite{DiatDataBase}. All errors shown below come from the fitting of the potential to a Morse potential~\footnote{ The only error accounted for is the error due to the fit of the potential to a Morse potential. This error is small, since the rotationally averaged potential $\widetilde{V(r)}$ fits very well to a Morse potential (both R-squared values are over 0.99). Errors due to the level of theory used or the limited size of the basis set used are not accounted for and are likely the most significant sources of error.}.}\label{tab:results}
\begin{tabular}{cccccc}
\hline
Species & \begin{tabular}[c]{@{}c@{}}Re\\ (\AA)\end{tabular} & \begin{tabular}[c]{@{}c@{}}$\omega_e$\\ (cm$^{-1}$)\end{tabular} & \begin{tabular}[c]{@{}c@{}}$\omega_e \chi_e$\\ (cm$^{-1}$)\end{tabular} & \begin{tabular}[c]{@{}c@{}}$Be$\\ (cm$^{-1}$)\end{tabular} & \begin{tabular}[c]{@{}c@{}}$\alpha_e$\\ (cm$^{-1}$)\end{tabular} \\ \hline
N$_2$@C$_{60}$ & 1.11 & 2376$\pm$4 & 12.30$\pm$0.03 & 1.95 & 0.01455$\pm$5$\times$10$^{-5}$ \\ \hline
N$_2$ & 1.097685 & 2358.57 & 14.324 & 1.998241 & 0.017318 \\ \hline
AlF@C$_{60}$ & 1.69 & 843$\pm$3 & 3.18$\pm$0.02 & 0.53 & 0.00288$\pm$10$^{-5}$ \\ \hline
AlF & 1.654369 & 802.26 & 4.77 & 0.55248 & 0.00498 \\ \hline \hline
\end{tabular}
\end{center}
\end{table*}

In a Morse potential, the spectroscopic constants of the molecule depend uniquely on the reduced mass of the molecule, $\mu$, and the three parameters of the Morse potential: $R_e$, $D_e$ and $a$. For instance, the vibrational harmonic frequency is given by~\cite{Herzberg}:
\begin{equation}
    \omega_e = \frac{a}{2\pi c}\sqrt{\frac{2D_e}{\mu}},
\end{equation}
where $c$ is the speed of light. The first anharmonic correction $\omega_e \chi_e$ is then given as:
\begin{equation}
    \omega_e \chi_e = \frac{\hbar a^2}{4\pi \mu c},
\end{equation}
where $\hbar$ is the reduced Planck's constant. The rotational constant $B_e$ is calculated directly from the equilibrium distance as
\begin{equation}
    B_e = \frac{\hbar}{4 \pi c \mu R_e^2}.
\end{equation}
Finally, the first correction to the rotational constant is:
\begin{equation}
    \alpha_e = \frac{3 \hbar^2 \omega_e}{4\mu R_e^2 D_e}(\frac{1}{aR_e}-\frac{1}{a^2R_e^2}).
\end{equation}
The resulting spectroscopic constants are shown in Table:~\ref{tab:results} for N$_2$@C$_{60}$ and AlF@C$_{60}$. The errors shown in Table:~\ref{tab:results} come from errors in $D_e$ and $a$ resulting from the fit. These errors are then propagated through the equations above. Overall, the spectroscopic constants of a free molecule and the molecule in a fullerene are very similar. In other words, the intrinsic properties of a molecule in the gas phase and a molecule inside a fullerene are roughly the same. However, there are general trends worth discussing. For instance, molecules appear stretched inside the fullerene with respect to their free case. As a result, the rotational constant of a molecule inside a fullerene is smaller than the rotational constant when the molecule is in free space. In a similar vein, $\omega_e$ is slightly larger for molecules in fullerenes than it is for the same molecules in free space.




\section{Dipole moment}
Heteronuclear molecules show a permanent dipole moment, which can be viewed as another spectroscopic constant and, as such, may be affected by the presence of the cage when the molecule is placed in a fullerene. Studies of HF@C$_{60}$ have shown that the fullerene cage can shield the dipole moment of the internal molecule~\cite{HF@C60}.

Here, we focus on AlF@C$_{60}$. To calculate its dipole moment, we use the B3LYP-D3 functional and the Def2TZpp basis set including the counterpoise correction. Using this approach, we obtained a dipole moment for free AlF of 1.51 Debye, which agrees with the experimentally determined value of 1.530 Debye \cite{NistDiatomic}. For AlF@C$_{60}$, we find a dipole moment of 1.63 Debye. Therefore, it seems that the cage slightly enhances the dipole moment of the internal molecule. This is the opposite of what is seen in HF@C$_{60}$\cite{HF@C60}, where the fullerene cage has been measured to suppress the dipole moment of the internal HF molecule by as much as 75\%. However, as noted above, the molecule is stretched when placed inside the fullerene, so if the Pauling ionic character is similar, then the dipole moment of the molecule inside the cage should be larger than in the free case. In addition, it could be possible that the influence of the fullerene on the dipole moment of the internal molecule changes dramatically depending on the type of internal molecule. Polarizabilities of endofullerenes have been shown to be different for different types of internal molecules \cite{sabirov2014polarizability}, so it is reasonable for the effects on the dipole moment to depend on the internal molecule as well. There is also the possibility that the discrepancy between our findings and the results for HF@C$_{60}$ is a failure of the quantum chemistry methods implemented. This could also be a result of temperature which is not accounted for in our study but is accounted for in the study of HF@C$_{60}$. In any case, this contradiction warrants future study.

\section{Conclusions}
After exploring the interaction energy landscape of AlF@C$_{60}$ and N$_2$@C$_{60}$, we predict that the spectroscopic constants of an internal molecule of an endofullerene are not significantly altered by the interaction with the carbon cage. We report no significant changes to the equilibrium distance, the rotational constant, the first correction to the rotational constant, the harmonic frequency and the first anharmonic correction. This agreement does provide some evidence that our approach is accurate. While the properties of the internal molecule do seem to remain mostly unchanged, the molecule-cage interaction is highly anisotropic, inducing strong interactions between the rotational states of the molecule. This strong interaction causes the molecule to have a high degree of alignment with the C$_{60}$ cage. In the case of an internal heteronuclear molecule, the molecule is predicted to have both strong alignment and strong orientation with the fullerene. 

Our results show a slight enhancement of the dipole moment of AlF when placed inside a C$_{60}$ molecule. Measurements with HF@C$_{60}$~\cite{HF@C60} reveal the opposite result. In HF@C$_{60}$, a strong suppression of the dipole moment of HF is observed. Studies of endofullerenes with heteronuclear diatomic molecules are scarce. A better understanding of this disagreement is a topic for future work along with calculations of a full potential surface for N$_2$ or AlF inside of a fullerene.

Based these results, we propose using off-resonant light, with respect to the fullerene transitions, to create laser-induced alignment of the internal molecule of an endofullerene. For instance, using a short laser pulse (pulse durations much shorter than the rotational period of the internal molecule) will lead to the typical time-revivals on the alignment of the molecule~\cite{Revivals,Rabitz,Revivals2,Stapelfeldt2003}. In this scenario, the revival times should be very similar between a free molecule and one inside a fullerene cage, since these revival times only depend on the rotational constant of the molecule. Therefore, any deviation between a free molecule and a molecule in a fullerene cage can be used to establish the properties of the endofullerene. Perhaps more importantly, this proposed experiment could be used to study the internal molecule's interaction with the fullerene.

\section{Acknowledgments}

The authors acknowledge support from the Simons Foundation.

\bibliography{apssamp}

\end{document}